# MODERN QUANTUM TECHNOLOGIES OF INFORMATION SECURITY

Oleksandr **Korchenko**[1], Yevhen **Vasiliu**[2], Sergiy **Gnatyuk**[3]

[1,3]Dept of Information Security Technologies, National Aviation University,
Kosmonavta Komarova Ave 1, 03680 Kyiv, Ukraine
[2]Dept of Information Technologies and Control Systems, Odesa National Academy of Telecommunications n.a. O.S. Popov,
Koval`ska Str 1, 65029 Odesa, Ukraine
E-mails: [1]icaocentre@nau.edu.ua, [2]vasiliu@ua.fm, [3]s.gnatyuk@nau.edu.ua

In this paper, the systematisation and classification of modern quantum technologies of information security against cyber-terrorist attack are carried out. The characteristic of the basic directions of quantum cryptography from the viewpoint of the quantum technologies used is given. A qualitative analysis of the advantages and disadvantages of concrete quantum protocols is made. The current status of the problem of practical quantum cryptography use in telecommunication networks is considered. In particular, a short review of existing commercial systems of quantum key distribution is given.

## 1. Introduction

Today there is virtually no area where information technology (IT) is not used in some way. Computers support banking systems, control the work of nuclear reactors, and control aircraft, satellites and spacecraft. The high level of automation therefore depends on the security level of IT. The latest achievements in communication systems are now applied in aviation. These achievements are public switched telephone network (PSTN), circuit switched public data network (CSPDN), packet switched public data network (PSPDN), local area network (LAN), and integrated services digital network (ISDN) [73]. These technologies provide data transmission systems of various types: surface-to-surface, surface-to-air, air-to-air, and space telecommunication. Cyber-terrorist attacks [78] can cause economic damage to aircraft companies and can also reduce flight security or cause casualties. Protection against such attacks is therefore an important scientific and technical problem.

One of the most effective ways of ensuring confidentiality and data integrity during transmission is cryptographic systems. The purpose of such systems is to provide key distribution, authentication, legitimate users authorisation, and encryption. Key distribution is one of the most important problems of cryptography. This problem can be solved with the help of [58]: classical information-theoretic schemes (requires channel with noise; efficiency is very low, 1–5%), classical public-key cryptography schemes (Diffie-Hellman scheme, digital envelope scheme; it has computational security), classical computationally secure symmetric-key cryptographic schemes (requires a pre-installed key on both sides and can be used only as scheme for increase in key size but not as key distribution scheme), quantum key distribution (provides information-theoretic security; it can also be used as a scheme for increase in key length), Trusted Couriers Key Distribution (it has a high price and is dependent on the human factor).

In recent years, quantum cryptography (QC) has attracted considerable interest. Quantum key distribution (QKD) [2–5, 9–12, 16, 22, 25, 26, 29, 30, 32, 39, 41–43, 45, 48, 51, 55–58, 61, 71, 74, 78] plays a dominant role in QC. The overwhelming majority of theoretic and practical research projects in QC are related to the development of QKD protocols. The number of different quantum technologies of information security (QTIS) is increasing, but there is no information about classification of these technologies in scientific literature (there are only a few works concerning classification of QKD protocols, for example [30, 57]). This makes it difficult to estimate the level of the latest achievements and does not allow using QTIS with full efficiency. *The purpose of this article* is the systematisation and classification of up-to-date quantum technologies of data (transmitted via telecommunication channels) security, analysis of their strengths and weaknesses, and prospects and difficulties of implementation. Quantum technologies of information security consist of quantum key distribution, quantum secure direct communication [7, 8, 13, 14, 17, 62, 70, 74–76], quantum secret sharing [21, 34, 44, 52, 67, 69], quantum stream cipher [1, 19, 35, 36, 47, 68], quantum digital signature [33, 63, 66], quantum steganography [18, 20, 40], etc.

## 2. Quantum key distribution

QKD includes the following protocols: *protocols using single (non-entangled) qubits (two-level quantum systems) and qudits* ($d$-level quantum systems, $d>2$) [2–4, 9–12, 16, 29, 30, 32, 39, 43, 45, 51, 55–58, 61, 71, 78]; *protocols using phase coding* [4, 30]; and *protocols using entangled states* [25, 26, 41, 42].

The main task of QKD protocols is encryption key generation and distribution between two users connecting via quantum and classical channels [30]. In 1984 Ch. Bennet from IBM and G. Brassard from Montreal University introduced the first QKD protocol [2, 22, 57, 58], which has become an alternative solution for the problem of key distribution. This protocol is called BB84 [9, 57, 78] and it refers to QKD protocols using single qubits. The states of these qubits are the polarisation states of single photons. The BB84 protocol uses four polarisation states of photons ($0°$, $45°$, $90°$, $135°$). These states refer to two mutually unbiased bases [48]. Error searching and correcting is performed using classical public channel, which need not be confidential but only authenticated. For the detection of intruder actions in the BB84 protocol, an error control procedure is used, and for providing unconditionally security a privacy amplification procedure is used [3]. The efficiency of the BB84 protocol equals 50%. Efficiency means the ratio of the photons number which are used for key generation to the general number of transmitted photons. Six-state protocol requires the usage of four states, which are the same as in the BB84 protocol, and two additional directions of polarization: right circular and left circular [12]. Such changes decrease the amount of information, which can be intercepted. But on the other hand, the efficiency of the protocol decreases to 33%. Next, the 4+2 protocol is intermediate between the BB84 and B92 protocol [39]. There are four different states used in this protocol for encryption: 0 and 1 in two bases. States in each bases are selected non-orthogonal. Moreover, states in different bases must also be pairwise non-orthogonal. This protocol has a higher IS level than the BB84 protocol, when weak coherent pulses but not a single photon source are used by sender [39]. But the efficiency of the 4+2 protocol is lower than efficiency of BB84 protocol. In the Goldenberg-Vaidman protocol [32], encryption of 0 and 1 is performed using two orthogonal states. Each of these two states is the superposition of two localised normalised wave packets. For protection against intercept-resend attack, packets are sent at random times. A modified type of Goldenberg-Vaidman protocol is called the Koashi-Imoto protocol [43]. This protocol does not use a random time for sending packets, but it uses an interferometer's non-symmetrisation (the light is broken in equal proportions between both long and short interferometer arms).

The measure of QKD protocol security is Shannon's mutual information between legitimate users (Alice and Bob) and an eavesdropper (Eve): $I_{AE}(D)$ and $I_{BE}(D)$, where $D$ is error level which is created by eavesdropping. For most attacks on QKD protocols, $I_{AE}(D) = I_{BE}(D)$, we will therefore use $I_{AE}(D)$. The lower $I_{AE}(D)$ in the extended range of $D$ is, the more secure the protocol is.

Six-state protocol and BB84 protocol were generalised in case of using $d$-level quantum systems—qudits instead qubits [16]. This allows increasing the information capacity of protocols. We can transfer information using $d$-level quantum systems (which correspond to the usage of trits, quarts, etc.) unlike the classical transmission systems, which use bits. It is important to notice that QKD protocols are intended for classical information (key) transfer via quantum channel.

The generalisation of BB84 protocol for qudits is called protocol using single qudits and two bases due to use of two mutually unbiased bases for the eavesdropping detection. Similarly, the generalisation of six-state protocol is called protocol using qudits and $d+1$ bases. These protocols' security against intercept-resend attack and non-coherent attack was investigated in a number of articles (see e.g. [16]). In [61] comparative analysis of the efficiency and security of different protocols using qudits (on the basis of known formulas for mutual information) are carried out.

In fig. 1 dependences of $I_{AB}(D)$, $I_{AE}^{(d+1)}(D)$ and $I_{AE}^{(2)}(D)$ are presented, where $I_{AB}(D)$ is mutual information between Alice and Bob and $I_{AE}^{(d+1)}(D)$ and $I_{AE}^{(2)}(D)$ is mutual information between Alice and Eve for protocols using $d+1$ and two bases accordingly.

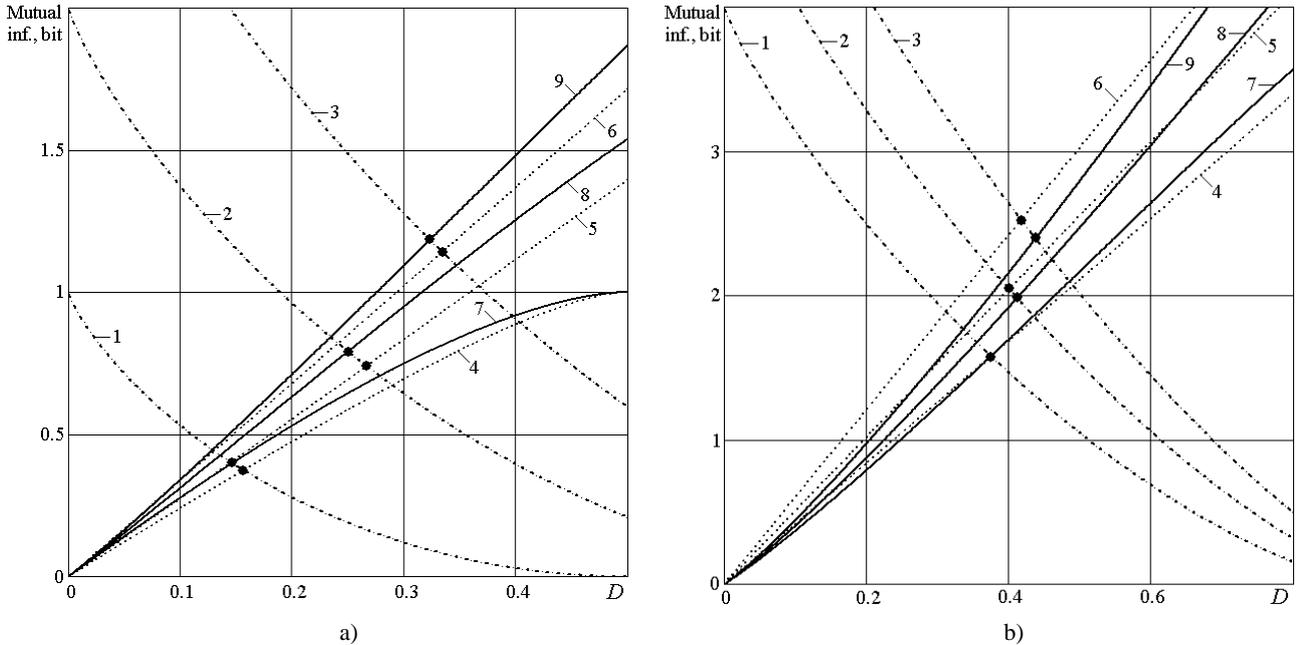

**Fig. 1.** Mutual information for non-coherent attack. 1, 2, 3— $I_{AB}(D)$ for $d$ = 2, 4, 8 (a) and $d$ = 16, 32, 64 (b); 4, 5, 6— $I_{AE}^{(d+1)}(D)$ for $d$ = 2, 4, 8 (a) and $d$ = 16, 32, 64 (b); 7, 8, 9— $I_{AE}^{(2)}(D)$ (6) for $d$ = 2, 4, 8 (a) and $d$ = 16, 32, 64 (b).

In fig. 1 we can see that at low qudit dimension (up to d ~ 16) the protocol's security against non-coherent attack is higher when $d+1$ bases are used (when $d$ = 2 it corresponds as noted above to greater security of six-state protocol than

BB84 protocol). But the protocol's security is higher when two bases are used in the case of large *d*, while the difference in Eve's information (using *d*+1 or two bases) is not large in the work region of the protocol, i.e. in the region of Alice's and Bob's low error level. We can conclude that the number of bases used has little influence on the security of the protocol against non-coherent attack (at least for the qudit dimension up to *d* = 64). The crossing points of curves $I_{AB}(D)$ and $I_{AE}(D)$ correspond to boundary values *D*, up to which one's legitimate users can establish a secret key by means of a privacy amplification procedure (even when eavesdropping occurs) [3].

Article [61] shows that the security of a protocol with qudits using two bases against intercept-resend attack is practically equal to the security of this protocol against non-coherent attack at any *d*. At the same time, the security of the protocol using *d*+1 bases against this attack is much higher. Intercept-resend attack is the weakest of all possible attacks on QKD protocols, but on the other hand, the efficiency of the protocol using *d*+1 bases rapidly decreases as *d* increases. A protocol with qudits using two bases therefore has higher security and efficiency than a protocol using *d*+1 bases.

Another type of QKD protocol is a protocol using phase coding [30]: for example, the B92 protocol [4] using strong reference pulses [30]. An eavesdropper can obtain more information about the encryption key in the B92 protocol than in the BB84 protocol for the given error level, however. Thus, the security of the B92 protocol is lower than the security of the BB84 protocol [29]. The efficiency of the B92 protocol is 25%.

The Ekert protocol (E91) [26, 30, 41] refers to QKD protocols using entangled states. Entangled pairs of qubits that are in a singlet state $|\psi^-\rangle = 1/\sqrt{2}(|0\rangle|1\rangle - |1\rangle|0\rangle)$ are used in this protocol. Qubit interception between Alice to Bob does not give Eve any information because no coded information is there. Information appears only after legitimate users make measurements and communicate via classical public authenticated channel [26]. But attacks with additional quantum systems (ancillas) are nevertheless possible on this protocol [41].

In article [42] generalisation of the Ekert scheme for three-level quantum systems introduced and in [25] generalisation of the Ekert scheme for *d*-level quantum systems is proposed: this increases the information capacity of the protocol a lot. Also in [25] the security of the protocol using entangled qudits is investigated. In article [61], based on the results of [25], the security comparison of protocol using entangled qudits and protocols using single qudits [16] against non-coherent attack is made. It was found that the security of these two kinds of protocols is almost identical. But the efficiency of the protocol using entangled qudits increases more slowly with the increasing dimension of qudits than the efficiency of the protocol using single qudits and two bases. Thus, from all contemporary QKD protocols using qudits, the most effective and secure against non-coherent attack is the protocol using single qudits and two bases (BB84 for qubits).

The aforementioned protocols with qubits are vulnerable to photon number splitting attack. This attack cannot be applied when the photon source emits exactly one photon. But there are still no such photon sources. Therefore, sources with Poisson distribution of photon number are used in practice. The part of pulses of this source has more than one photon. That is why Eve can intercept one photon from pulse (which contains two or more photons) and store it in quantum memory until Alice transfers Bob the sequence of bases used. Then Eve can measure stored states in correct basis and get the cryptographic key while remaining invisible. It should be noted that there are more advanced strategies of photon number splitting attack which allow Bob to get the correct statistics of the photon number in pulses if Bob is controlling these statistics [45]. In practice for realisation of BB84 and six-state protocols weak coherent pulses with average photon number about 0.1 are used. This allows avoiding small probability of two- and multi-photon pulses, but this also considerably reduces the key rate.

The SARG04 protocol does not differ much from the original BB84 protocol [10, 56, 57]. The main difference does not refer to the 'quantum' part of the protocol; it refers to the 'classical' procedure of key sifting, which goes after quantum transfer. Such improvement allows increasing security against photon number splitting attack. The SARG04 protocol in practice has a higher key rate than the BB84 protocol [10].

Another way of protecting against photon number splitting attack is the use of decoy states QKD protocols [11, 51, 55, 57, 71], which are also advanced types of BB84 protocol. In such protocols, besides information signals Alice's source also emits additional pulses (decoys) in which the average photon number differs from the average photon number in the information signal. Eve's attack will modify the statistical characteristics of the decoy states and/or signal state and will be detected. As practical experiments have shown for these protocols (as for the SARG04 protocol), the key rate and practical length of the channel is bigger than for BB84 protocols [51, 55, 71]. Nevertheless, it is necessary to notice that using these protocols, as well as the others considered above, it is also impossible without users pre-authentication to construct the complete high-grade solution of the problem of key distribution.

As a conclusion, after the analysis of the first and scale QTIS method, we must sum up and highlight the following advantages of QKD protocols: 1) These protocols always allow eavesdropping to be detected because Eve's connection brings much more error level (compared with natural error level) to the quantum channel. The laws of quantum mechanics allow eavesdropping to be detected and the dependence between error level and intercepted information to be set. This allows applying privacy amplification procedure, which decreases the quantity of information about the key, which can be intercepted by Eve. Thus, QKD protocols have unconditional (information-theoretic) security; 2) the information-theoretic security of QKD allows using an absolutely secret key for further encryption using well-known classical symmetrical algorithms. Thus, the entire information security level increases. It is also possible to synthesize QKD protocols with Vernam cipher (one-time pad) which in complex with unconditionally secured authenticated schemes gives a totally secured system for transferring information.

The disadvantages of quantum key distribution protocols are: 1) A system based only on QKD protocols cannot serve as a complete solution for key distribution in open networks (additional tools for authentication are needed); 2) the limitation of quantum channel length which is caused by the fact that there is no possibility of amplification without quantum properties being lost; 3) need for using weak coherent pulses instead of single photon pulses. This decreases the efficiency of protocol in practice. But this technology limitation might be defeated in the nearest future; 4) the data transfer rate decreases rapidly with the increase in the channel length. When the channel length is 100 km, the data transfer rate equals few bps; 5) photon registration problem which leads to key rate decreasing in practice; 6) photon depolarization in the quantum channel. This leads to errors during data transfer. Now the typical error level equals a few percent, which is much greater than the error level in classical communication systems; 7) difficulty of the practical realisation of QKD protocols for $d$-level quantum systems; 8) the high price of commercial QKD systems.

## 3. Quantum secure direct communication

The next method of information security based on quantum technologies is the usage of quantum secure direct communication (QSDC) protocols [7, 8, 13, 14, 17, 62, 70, 74–76]. The main feature of QSDC protocols is that there are no cryptographic transformations; thus, there is no key distribution problem in QSDC. In these protocols, a secret message is coded by qubits' (qudits') quantum states, which are sent via quantum channel. QSDC protocols can be divided into several types: *ping-pong protocol (and its enhanced variants)* [7, 14, 75, 76], *protocols using block transfer of entangled qubits* [17, 62], *protocols using single qubits* [13] and *protocols using entangled qudits* [62]. There are QSDC protocols for two parties and for multi-parties, e.g. broadcasting or when one user sends message to another under the control of a trusted third party.

Most contemporary protocols require a transfer of qubits by blocks [17, 62]. This allows eavesdropping to be detected in the quantum channel before transfer of information. Thus, transfer will be terminated and Eve will not obtain any secret information. But for storing such blocks of qubits there is a need for a large amount of quantum memory. The technology of quantum memory is actively being researched, but it is still far from usage in common standard telecommunication equipment. So from the viewpoint of technical realisation, protocols using single qubits or their non-large groups (for one cycle of protocol) have an advantage. There are few such protocols and they have only asymptotic security, i.e. the attack will be detected with high probability, but Eve can obtain some part of information before detection. Thus, the problem of privacy amplification appears. In other words, new pre-processing methods of transferring information are needed. Such methods should make intercepted information negligible.

One of the quantum secure direct communication protocols is the ping-pong protocol [7, 14, 62, 75, 76], which does not require qubit transfer by blocks. In the first variant of this protocol, entangled pairs of qubits and two coding operations that allow the transmission of one bit of classical information for one cycle of the protocol are used [7]. The usage of quantum superdense coding allows transmitting two bits for a cycle [14]. The subsequent increase in the informational capacity of the protocol is possible by the usage instead of entangled pairs of qubits their triplets, quadruplets etc. in Greenberger-Horne-Zeilinger (GHZ) states [76]. The informational capacity of the ping-pong protocol with GHZ-states is equal to $n$ bits on a cycle where $n$ is the number of entangled qubits. Another way of increasing the informational capacity of ping-pong protocol is using entangled states of qudits. Thus, the corresponding protocol based on Bell's states of three-level quantum system (qutrit) pairs and superdense coding for qutrits is introduced in [62, 75].

The advantages of QSDC protocols are a lack of secret key distribution, the possibility of data transfer between more than two parties, and the possibility of attack detection providing a high level of IS (up to information-theoretic security) for the protocols using block transfer. The main disadvantages are difficulty in practical realisation of protocols using entangled states (and especially protocols using entangled states for $d$-level quantum systems), slow transfer rate, the need for large capacity quantum memory for all parties (for protocols using block transfer of qubits), and the asymptotic security of the ping-pong protocol. Besides, QSDC protocols similarly to QKD protocols is vulnerable to man-in-the-middle attack, although such attack can be neutralized by using authentication of all messages, which are sent via the classical channel.

Asymptotic security of the ping-pong protocol (which is one of the simplest QSDC protocols from the technical viewpoint) can be amplified by using methods of classical cryptography. Security of several types of ping-pong protocols using qubits and qutrits against different attacks was investigated in series of works [7, 14, 70, 75, 76]. The security of the ping-pong protocol using qubits against eavesdropping attack using ancilla states is investigated [17, 76]. In fig. 2 dependences of composite probability of not detecting an attack for the ping-pong protocol with many-qubit GHZ-states are shown. It is obvious from fig. 2 that the ping-pong protocol with many-qubit GHZ-states is asymptotically secure at any number $n$ of qubits that are in entangled GHZ-states. A similar result for the ping-pong protocol using Bell states of qutrit pairs is presented [75].

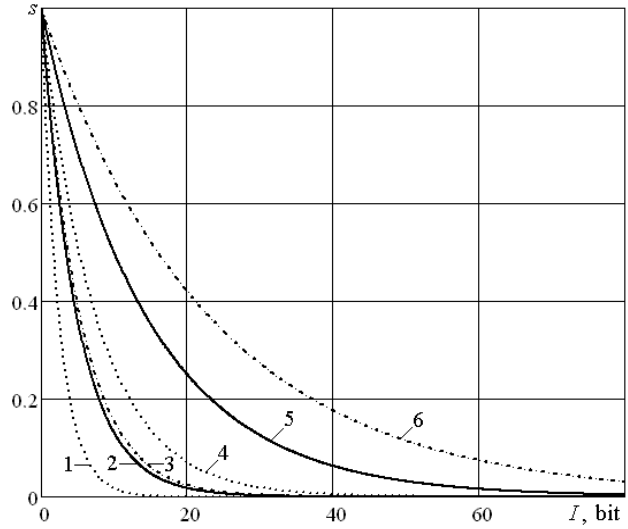

**Fig. 2.** Composite probability of attack non-detection $s$ for the ping-pong protocol with many-qubit GHZ-states: $n=2$, original protocol (1); $n=2$, with superdense coding (2); $n=3$ (3); $n=5$ (4); $n=10$ (5); $n=16$ (6). $I$ is Eve's information.

A non-quantum method of security amplification for the ping-pong protocol is suggested in [76]. This method is as follows. Before the transmission, Alice divides the binary message on $l$ block of some fixed length $r$; we will designate these blocks through $a_i$ ($i=1, \ldots, l$). Alice then generates for each block separately random invertible binary matrix $K_i$ of size $r \times r$ and multiplies these matrices by appropriate blocks of the message $b_i = K_i a_i$ (multiplication is performed by modulo 2). Blocks $b_i$ are transmitted on the quantum channel with the use of the ping-pong protocol. Even if Eve manages to intercept one (or more) from these blocks while remaining undetected, not knowing matrices $K_i$ used, Eve cannot reconstruct source blocks $a_i$. To reach sufficient security level, the block length $r$ and accordingly the size of matrices $K_i$ should be selected so that Eve's probability of non-detection $s$ after the transmission of one block is insignificant small. Matrices $K_i$ are transmitted to Bob via usual (non-quantum) open authentic channel after the end of quantum transmission but only in the event that Alice and Bob are convinced of lack of eavesdropping. Bob then inverses the received matrices and having multiplied them on appropriate blocks $b_i$ he gains the original message.

This method allows providing high security level of the ping-pong protocol (choosing suitable length of blocks for hashing). Rounded values of block length $r$ for the ping-pong protocol with $n$-qubit GHZ-states at $s = 10^{-6}$ and for the case when Eve aspires to get all information and makes maximal error level for legitimate users are presented in table 1. The probability of detecting the attack is maximal in this case [76]. The quantity of $q$ is a probability of switching to control mode [7, 76].

**Table 1.** Rounded values of block length $r$ for the ping-pong protocol with $n$-qubit GHZ-states (bit)

| $n$ | $q = 0.5$, $d = d_{max}$ | $q = 0.25$, $d = d_{max}$ |
|---|---|---|
| 2 | 69 | 180 |
| 3 | 74 | 186 |
| 4 | 88 | 216 |
| 5 | 105 | 254 |
| 6 | 123 | 297 |
| 7 | 142 | 341 |
| 8 | 161 | 387 |
| 9 | 180 | 434 |
| 10 | 200 | 481 |
| 11 | 220 | 529 |
| 12 | 240 | 577 |
| 13 | 260 | 625 |
| 14 | 279 | 673 |
| 15 | 299 | 721 |

| 16 | 319 | 769 |
|----|-----|-----|
| 17 | 339 | 817 |
| 18 | 359 | 865 |
| 19 | 379 | 913 |
| 20 | 399 | 961 |

Thus, after transfer of hashed block, the lengths of which are presented in tab. 1, the probability of attack non-detection will be equal to $10^{-6}$; there is thus a very high probability that this attack will be detected. The main disadvantage of the ping-pong protocol, namely its asymptotic security against eavesdropping attack using ancilla states, is therefore removed. There are some others attacks on the ping-pong protocol, e.g. attack which can be performed when the protocol is executed in quantum channel with noise [70]. But there are some counteraction methods to these attacks [8]. Thus, we can say that the ping-pong protocol (the security of which is amplified using method described above) is the most prospective QSDC protocol from the viewpoint of the existing development level of the quantum technology of information processing.

**4. Other quantum methods of information security**

*Quantum secret sharing (QSS).* Most QSS protocols use properties of entangled states [9, 48]. The first QSS protocol was proposed by Hillery, Buzek and Berthiaume in 1998 [34, 52]. This protocol uses GHZ-triplets (quadruplets) similar to some QSDC protocols. The sender shares his message between two (three) parties and only cooperation allows them to read this message. Semi-quantum secret sharing protocol using GHZ-triplets (quadruplets) is proposed in [44]. In this protocol, users that receive a shared message have access to the quantum channel. But they are limited by some set of operation and are called 'classical', meaning they are not able to prepare entangled states and perform any quantum operations or measurements. These users can measure qubits on a 'classical' $\{|0\rangle, |1\rangle\}$ basis, reordering the qubits (via proper delay measurements), preparing (fresh) qubits in the classical basis, and sending or returning the qubits without disturbance. The sending party can perform any quantum operations. This protocol prevails over others QSS protocols in economic terms. Its equipment is cheaper because expensive devices for preparing and measuring (in GHZ-basis) many-qubit entangled states are not required. Semi-quantum secret sharing protocol exists in two variants: randomisation-based and measurement-resend protocols. In article [69] QSS using single qubits that are prepared in two mutually unbiased bases and transferred by blocks is presented. Similar to the Hillery-Buzek-Berthiaume protocol, this allows sharing a message between two (or more) parties. The security improvement of this protocol against malicious acts of legitimate users is proposed in [21]. A similar protocol for multiparty secret sharing is presented in article [67]. QSS protocols are protected against external attackers and unfair actions of the protocol's parties. Both quantum and semi-quantum schemes allow detecting eavesdropping and do not require encryption unlike the classical secret-sharing schemes. The most significant imperfection of QSS protocols is the necessity for large quantum memory that is outside the capabilities of modern technologies today.

*Quantum stream cipher (QSC)* provides data encryption similar to classical stream cipher, but it uses quantum noise effect [36] and can be used in optical communication networks. QSC is based on the Yuen-2000 protocol (Y-00, $\alpha\eta$ - scheme). Information-theoretic security of the Y-00 protocol is ensured by randomisation (based on quantum noise) and additional computational schemes [68]. In articles [19, 35] the high encryption rate of the Y-00 protocol is demonstrated experimentally, and in [35] a security analysis on the Yuen-2000 protocol against the fast correlation attack, the typical attack on stream ciphers, is presented. The next advantage is better security compared with usual (classical) stream cipher. This is achieved by quantum noise effect and by the impossibility of cloning quantum states [65]. The complexity of practical implementation is the most important imperfection of QSC [35].

*Quantum digital signature (QDS)* can be implemented on the basis of protocols such as QDS protocols using single qubits [63] and QDS protocols using entangled states (authentic QDS based on quantum GHZ-correlations) [66]. QDS is based on use of the quantum one-way function [33]. This function has better security than the classical one-way function, and it has information-theoretic security (its security does not depend on the power of the attacker's equipment). Quantum one-way function is defined by the following properties of quantum systems [33]: 1) qubits can exist in superposition 0 and 1 unlike classical bits; 2) we can get only a limited quantity of classical information from quantum states (according to the Holevo theorem) [37, 48]. Calculation and validation are not difficult but inverse calculation is impossible. In the systems that use QDS, user identification and integrity of information is provided similar to classical digital signature [33]. The main advantages of QDS protocols are information-theoretic security and simplified key distribution system. The main disadvantage is the possibility to generate a limited number of public key copies and the leak of some quantities of information about incoming data of quantum one-way function (unlike the ideal classical one-way function) [33].

*Quantum steganography* [40] aims to hide the fact of information transferral similar to classical steganography. In articles [20, 40] models of quantum steganography systems are proposed, but there is no case of the practical implementation of these systems. All current models of quantum steganography systems use entangled states. For example, modified methods of entangled photon pair detection are used to hide the fact of information transfer in patent [18]. Theoretical research in this area has not reached the level of practical application yet, and it is very difficult to talk about the advantages and disadvantages of quantum steganography systems. Whether quantum steganography is superior to the classical one or not in practical use is still an open question [40].

Fig. 3 represents a general scheme of quantum methods of IS for their purposes and for using QTIS.

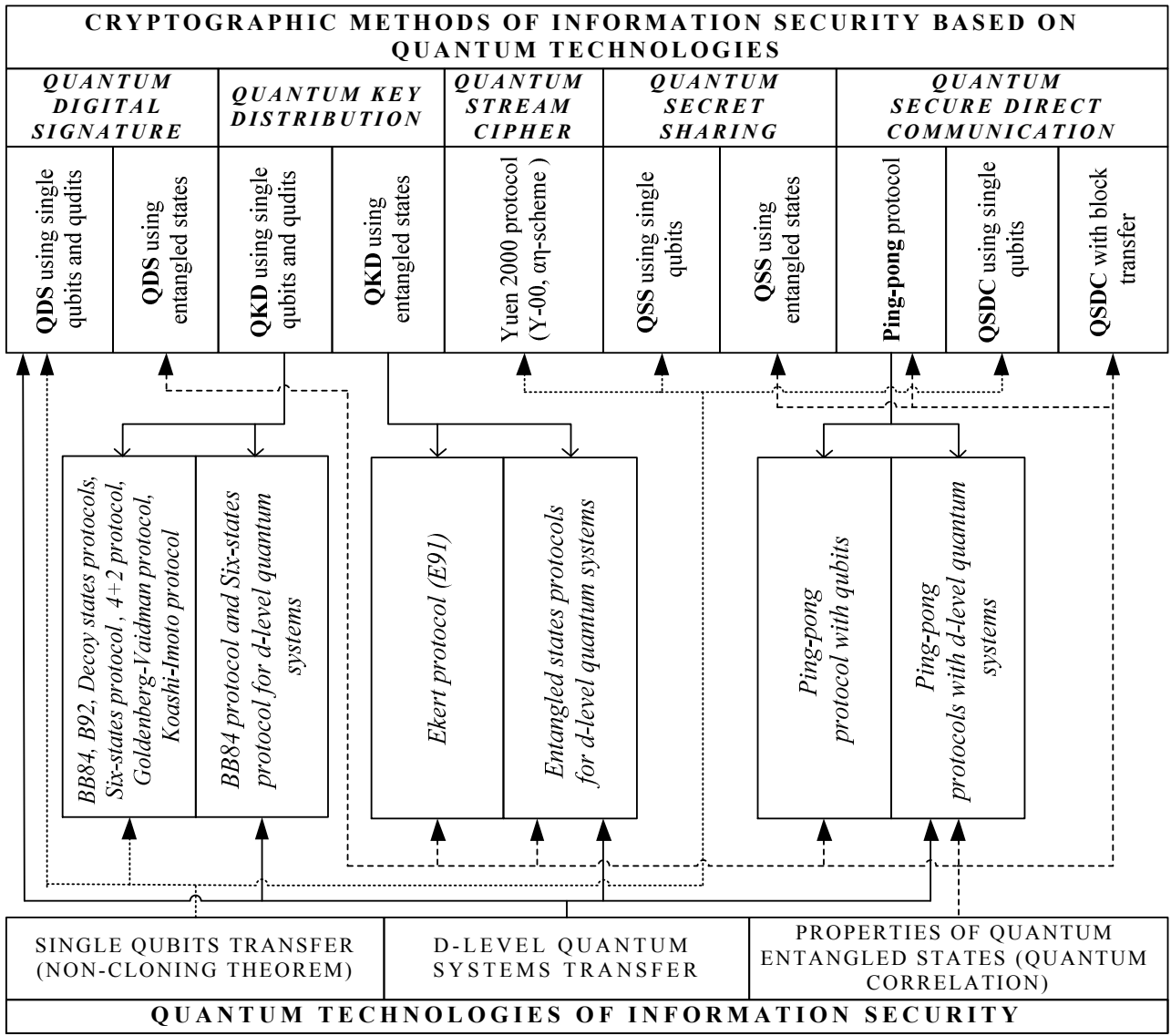

**Fig. 3.** Classification of quantum methods of IS

## 5. Commercial QKD Systems

The world's first commercial quantum cryptography solution was QPN Security Gateway (QPN-8505) [54] proposed by MagiQ Technologies (USA). This system is a cost-effective IS solution for governmental and financial organisations. It proposes VPN protection using QKD (up to 100 256-bit keys per second, up to 140 km) and integrated encryption. The QPN-8505 system uses BB84, 3DES [50] and AES [49] protocols. The Swiss company Id Quantique offers a system called Cerberis [15]. It is a server with automatic creation and secret key exchange over a fibre channel (FC-1G, FC-2G and FC-4G). This system can transmit cryptographic keys up to 50 km and carries out 12 parallel cryptographic calculations. The latter substantially improves the system's performance. The Cerberis system uses AES (256-bits) for encryption and BB84 and SARG04 protocols for quantum key distribution. Toshiba Research Europe Ltd (Great Britain) recently presented another QKD system named Quantum Key Server [53]. This system has a very simple architecture and provides up to 100 256-bit keys per second with their one-way transferring from sender to receiver. Quantum Key Server includes an integrated automatic control module that provides continuous monitoring and regulation of the system's optical characteristics. Another British company, QinetiQ, realised the world's first network using quantum cryptography—Quantum Net (Qnet) [27, 38]. The maximum length of communication lines in this network is 120 km. Moreover, it is a very important fact that Qnet is the first QKD system using more than two servers. This system has six servers integrated to the Internet.

In addition the world's leading scientists are actively taking part in the implementation of projects such as SECOQC (Secure Communication based on Quantum Cryptography) [58] and EQCSPOT (European Quantum Cryptography and Single Photon Technologies). There are many practical and theoretical research projects concerning the development of QTIS in research institutes, laboratories and centres (Northwestern University, BBN Technologies of Cambridge, TREL, NEC, Mitsubishi Electric, ARS Seibersdorf Research, Los Alamos National Laboratory) [72].

Most methods and facilities of quantum cryptography are patented [6, 23, 24, 28, 31, 46, 59, 60, 64, 77, 79, 80] in different countries and have the prospect to be realised in the near future.

## 6. Conclusions

This article presents a classification and systematisation of modern quantum technology of information security. The characteristic of the basic directions of quantum cryptography from the point of view of the quantum technologies used is given. A qualitative analysis of the advantages and imperfections of concrete quantum protocols is made. Today the most developed direction of quantum cryptography is QKD protocols. In research institutes, laboratories and centres, quantum cryptographic systems for secret key distribution for distant legitimate users are being developed. Most of the technologies used in these systems are patented in different countries (mainly in the U.S.A.). Such QKD systems can be combined with any classical cryptographic scheme, which provides information-theoretic security, and the entire cryptographic scheme will have information-theoretic security also. QKD protocols can generally provide higher IS level than appropriate classical schemes.

Other quantum technologies of information security (QTIS) in practice have not yet extended beyond laboratory experiments. But there are many theoretical cryptographic schemes that provide high IS level up to the information-theoretic security. Quantum secure direct communication protocols do not have any analogues in classical cryptography. These protocols remove the secret key distribution problem because they do not use encryption. One of these is the ping-pong protocol and its improved versions. These protocols can provide high IS level of confidential data transmission using the existing level of technology with security amplification methods. Another category of QSDC is protocols with transfer qubit by blocks that have unconditional security, but these need a large quantum memory which is outside the capabilities of modern technologies today. It should be noted that QSDC protocols are not suitable for the transfer of a high-speed flow of confidential data because there is low data transfer rate in the quantum channel. But when a high IS level is more important than transfer rate, QSDC protocols should find its application.

Quantum secret sharing protocols allow detecting eavesdropping and do not require data encryption. This is their main advantage over classical secret sharing schemes. Similarly, quantum stream cipher and quantum digital signature provide higher security level than classical schemes. Quantum digital signature has information-theoretic security because it uses quantum one-way function. However, practical implementation of these QTIS is also faced with some technological difficulties.

Thus, in recent years QTIS are rapidly developing and gradually taking their place among other means of IS. Their advantage is a high level of security and some properties, which classical means of IS do not have. One of these properties is the ability always to detect eavesdropping. QTIS therefore represent an important step towards improving the security of communication systems against cyber-terrorist attacks. But many theoretical and practical problems must be solved for practical the use of QTIS in existing communication systems.

Information about authors:

**Oleksandr KORCHENKO,** Prof Dr Habil.
*Date and place of birth:* 1961, Kyiv, Ukraine.
*Education:* Kiev Institute of Civil Aviation Engineers (National Aviation University since 2000), 1983.
*Affiliation and functions:* Dr Habil from National Aviation University since 2004, 2005 – professor at National Aviation University, head of Dept. of Information Technologies Security of National Aviation University since 2004.
*Research interests:* information and aviation security.
*Publications:* over 160 books and articles, 8 patents.

**Yevhen VASILIU,** Candidate of Science (PhD).
*Date and place of birth:* 1966, Yalta, Crimea, Ukraine.
*Education:* Odessa State University n.a. I.I. Mechnikov, 1990.
*Affiliation and functions:* PhD in theoretical physics since 1999, docent at Odesa National Academy of Telecommunications n.a. O.S. Popov since 2004.
*Research interests:* quantum information, quantum cryptography.
*Publications:* over 50 books and articles.

**Sergiy GNATYUK,**
*Date and place of birth:* 1985, Netishyn, Khmelnytskyi Oblast, Ukraine.
*Education:* National Aviation University, 2007.
*Affiliation and functions:* post-graduate student at National Aviation University since 2007.
*Research interests:* information security, quantum cryptography.
*Publications:* over 20 papers, 2 patents.